\documentclass[prd,aps,twocolumn,preprintnumbers,amsmath,amssymb,superscriptaddress]{revtex4}

\pdfoutput=1

\usepackage{amsmath}
\usepackage{graphicx}
\usepackage{epsfig}
\usepackage{dcolumn}
\usepackage{bm}
\usepackage{amssymb}
\usepackage{latexsym}
\usepackage{epstopdf}
\usepackage{threeparttable}
\usepackage{booktabs}
\usepackage{tabularx}
\usepackage{mathrsfs}
\usepackage{hhline}
\usepackage{multirow}
\usepackage{color}
\usepackage[colorlinks,linkcolor=magenta,anchorcolor=blue,citecolor=blue]{hyperref}

\def\be{\begin{equation}}
\def\ee{\end{equation}}
\def\bea{\begin{eqnarray}}
\def\eea{\end{eqnarray}}

\begin{document}
\title{Towards degeneracy breaking of early universe models}

\author{Ze Luan}
\email{luanze@mails.ccnu.edu.cn}
\affiliation{Institute of Astrophysics, Central China Normal University, Wuhan 430079, China}
\affiliation{Key Laboratory of Particle Astrophysics, Institute of High Energy Physics, Chinese Academy of Sciences, Beijing 100086, China}

\author{Taotao Qiu}
\email{qiutt@mail.ccnu.edu.cn}
\affiliation{Institute of Astrophysics, Central China Normal University, Wuhan 430079, China}
\affiliation{Key Laboratory of Particle Astrophysics, Institute of High Energy Physics, Chinese Academy of Sciences, Beijing 100086, China}

\begin{abstract}
There are many possibilities of scenarios in the early universe, which can give rise to the same observational signals due to the degeneracy among each other, caused by equivalence under the conformal transformations. In order to break the degeneracy, in this paper we take into account the so-called ``frame-invariant variables" proposed by A. Ijjas and P. J. Steinhardt in \cite{Ijjas:2015zma}. We discuss how the different scenarios will distribute in different parametric space constructed from those variables, waiting for the judgement of real observations in the future. Several concrete models with explicit non-minimal coupling functions are also discussed.
\end{abstract}

\maketitle
\section{introduction}
There have been more and more models aiming at describing the early stage of our universe. Besides the standard cosmological model that claims our universe begin from a Big-Bang point and expands forever \cite{Alpher:1948ve}, there are also viewpoints that our universe may come from a contracting phase followed by a bounce \cite{Novello:2008ra}, or even several bounces (a.k.a. cyclic) \cite{Lehners:2008vx}. Therefore, in spite of difficulty, it is important to make clear which is the real appearance of the early universe.

Perhaps the best way to distinguish different scenarios is the constraints from the observations. The most powerful observations of the early universe come from those on CMB photons, which carries information about primordial perturbations. The current data tells us that these perturbations have to 1) be flat and scale-invariant in its scalar mode, 2) have small enough tensor-to-scalar ratio, and 3) have small enough non-Gaussianities \cite{Akrami:2018odb, Ade:2015ava}. These conditions have excluded many candidates, while they favor the slow-roll inflation model, which is then acknowledged by most people and has been viewed as new standard model.

However, along with the theoretical development, people find that not only inflation satisfies those conditions. For example, in the beginning of this century, Ref. \cite{Finelli:2001sr} (see also \cite{Wands:1998yp}) found that a contracting phase behaving like matter domination can also produce flat and scale-invariant scalar perturbations, which is actually dual to inflation. However, this matter contraction (or known as matter bounce, if followed by a bounce to expanding phase) will have large tensor-to-scalar ratio. Moreover, if one considers nonminimal coupling to gravity, he will get much more possibilities. Due to the invariance of the power spectrum and non-Gaussianities under conformal transformation, one can get almost arbitrary backgrounds of the early universe, as long as can be related to inflation/matter-contraction by such a transformation \cite{Piao:2011mq, Ijjas:2015zma, Qiu:2012ia}. Therefore, there will be ``degeneracy" among those backgrounds, for example, when put contracting scenarios with nonminimal coupling into their Einstein frame, they will behave like slow-roll inflation, or vice versa. Since we cannot know which frame the universe should be in at its early stage, we cannot determine which model is of the real world, and which is just the ``image" in the mirror of conformal transformation.

How can we break this degeneracy? Since CMB observation itself cannot solve the question, we may have to refer to other methods. In 2015, Ijjas and Steinhardt have provided a novel method to consider frame-independent quantities that are {\it different} for the real ``object" and its ``image" \cite{Ijjas:2015zma}. If we could measure such quantities, we should be able to distinguish each other. However, to do this, we need to obtain the constraints on the time dependence of both particle mass $m$ and the Planck mass $M_{Pl}$, in terms of $\dot m/m$ and $\dot M_{Pl}/M_{Pl}$, respectively. While we have constraints on both quantities for current value, those for early universe seems lacking. However, this does not mean that we couldn't make any effort towards this way. In this paper, by parameterizing those frame-independent quantities, we're going to obtain regions in the parameter space corresponding to those scenarios, in order to have them distinguished. If there are real constraints on these quantities from observations in the future, we should be able to tell which is real ``object". We also discuss some concrete models as specific examples.

The rest of the paper is organized as follows: in Sec. II we review several characteristic scenarios that could give rise to scale-invariant power spectrum, and how they relates to each other. Although inflation and matter contraction have different tensor-to-scalar ratio, since both scenarios are well-known and interesting to us, both will be included in our analysis. In Sec. III we introduce the frame-independent quantities and analyze how they can be used to break the degeneracy between those models. In Sec. IIIA we try to parametrize these quantities in terms of $\dot m/m$ and $\dot M_{Pl}/M_{Pl}$. According to this, in Sec. IIIB we try to obtain regions in the parameter space for different scenarios, and in Sec. IIIC we discuss about concrete examples. Sec. IV contains our final remarks.





\section{The early universe scenarios with scale-invariant power spectrum}

\subsection{slow-roll inflation}
Inflation \cite{Guth:1980zm, Linde:1981mu, Starobinsky:1980te} is the most well-known scenario to generate scale-invariant scalar and tensor spectrum, especially under the ``slow-roll" assumption. In inflation scenario, we consider a scalar field (inflaton) which is slow-rolling down along the potential. Assuming the field to be canonical for simplicity, the action of the scalar field is of the form:
\be
\label{actioninf}
S_{inf}=\int d^4x\sqrt{-g}\left[\frac{M_{Pl}^2R}{2}-\frac{1}{2}(\nabla\phi)^2-V(\phi)\right]~,
\ee
and the ``slow-roll" condition is defined as $\dot\phi^2\ll 2V$, according to which one gets from the Friedmann equation:
\be
H^2\simeq \frac{1}{3M_{Pl}^2}V~,~~~\epsilon_{inf}\equiv-\frac{\dot H}{H^2}=\frac{\dot\phi^2}{H^2}\ll 1~,
\ee
where $H$ is the Hubble parameter and $\epsilon_{inf}$ is the slow-roll parameter, and the equation of motion can also be simplified as
\be
3H\dot\phi+V_{\phi}\simeq 0~.
\ee

To do the perturbations, one first perturb the metric such that the line element is
\be
ds^2=-(1+\alpha)^2dt^2+2\partial_i\beta dt dx^i+a^2(t) e^{2\zeta\delta_{ij}+\gamma_{ij}} dx^idx^j~,
\ee
where $\alpha$, $\beta$ and $\zeta$ are scalar perturbation modes, and $\gamma_{ij}$ is the tensor perturbation mode. In addition one could also have the perturbation of the field $\delta\phi$, but it can be moved away when choosing the uniform-$\phi$ gauge. Therefore, under straightforward calculation one gets the perturbation action up to the second order:
\be
\label{actionpertinfS}
\delta_2S^{(S)}_{inf}=\int d^4x a^3 \epsilon_{inf}\left[\dot\zeta^2-\frac{(\partial\zeta)^2}{a^2}\right]~,
\ee
which gives the equation of motion:
\be
\label{eompertinfS}
u^{\prime\prime}+k^2u-\frac{z^{\prime\prime}}{z}u=0~,
\ee
where $u\equiv z\zeta$, $z\equiv a(t)\sqrt{2\epsilon_{inf}}$, prime denotes derivative with respect to conformal time: $\tau=\int a^{-1}(t)dt$. From Friedmann equation one has
\be
a\sim t^{1/\epsilon_{inf}}\sim\tau^{1/(\epsilon_{inf}-1)}~,
\ee
and this leads to $\frac{z''}{z}\rightarrow\frac{2-\epsilon_{inf}}{(\epsilon_{inf}-1)^2\tau^2}$, which gives the solution
\bea
\label{solpertinfS}
u&=&c^{inf}_1\sqrt{|\tau|}H_{\nu}(|k\tau|)~,~c^{inf}_2\sqrt{|\tau|}H_{-\nu}(|k\tau|)~,\\
\nu&=&\frac{1}{\epsilon_{inf}-1}-\frac{1}{2}~,
\eea
where $H_\nu$ is the Hankel function of $\nu$-th order. In the superhorizon ($|k\tau|\rightarrow 0$) limit, one has $H_\nu(|k\tau|)\rightarrow\sqrt{2/\pi}(|k\tau|)^\nu$. Moreover, from the simple relation of $a$ and $\tau$ we also have another relationship: $aH=-1/\tau$. One then has
\be
\label{zetainf}
\zeta=\frac{u}{z}=\frac{c^{inf}_1H}{\sqrt{\pi\epsilon_{inf}}}k^\nu|\tau|^{\frac{3}{2}+\nu}~,~\frac{c^{inf}_2H}{\sqrt{\pi\epsilon_{inf}}}k^{-\nu}|\tau|^{\frac{3}{2}-\nu}~.
\ee
For the case of inflation, since $\epsilon\simeq 0$ and thus $\nu\simeq -3/2$, one can see the second mode of Eq. (\ref{zetainf}) is decaying and the  first mode (constant \footnote{The constant is exact, since $H$ also has a dependence of $\tau$, i.e. $H\sim \tau^{-\epsilon_{inf}/(\epsilon_{inf}-1)}$, which conpensate the deviation of $\nu$ from $-3/2$.}) is then dominant. Moreover, the initial condition of $u$ can be obtained from Eq. (\ref{eompertinfS}) by considering the large $k$ region, where the second term is much larger than the third term:
\be
\label{uini}
u_i=\frac{1}{\sqrt{2k}}e^{ik\tau}~,
\ee
which can be combined with Eq. (\ref{solpertinfS}) to get: $c^{inf}_1=\sqrt{\pi}/2$. The power spectrum and its spectral index of the scalar perturbation are then
\bea
\label{spectruminfS}
P^{S}_{inf}&\equiv&\frac{k^3}{2\pi^2}|\zeta|^2=\frac{H^2}{8\pi^2\epsilon_{inf}}|k\tau|^{3+2\nu}~,\\
n^{S}_{inf}&\equiv&1+\frac{d\ln P^{S}_{inf}}{d\ln k}\simeq 1-2\epsilon_{inf}~.
\eea
from which we can see that the scalar perturbation is nearly scale-invariant.

Similarly, one can also calculate the spectrum of tensor perturbation and its scale-dependence. The action of tensor perturbation up to the second order is:
\be
\label{actioninfT}
\delta_2S^{(T)}_{inf}=\int d^4x \frac{a^3}{4}\sum_{s=+,\times}\left[\dot\gamma^{(s)2}-\frac{(\partial\gamma^{(s)})^2}{a^2}\right]~,
\ee
where $\gamma^{(s)}$ is the nonzero component of $\gamma_{ij}$. Action (\ref{actioninfT}) gives the equation of motion:
\be
\label{eompertinfT}
v^{\prime\prime}+k^2v-\frac{a^{\prime\prime}}{a}v=0~,
\ee
where $v\equiv a\gamma/\sqrt{2}$. Moreover, $a\sim\tau^{1/(\epsilon_{inf}-1)}$ leads to $\frac{a''}{a}\rightarrow\frac{2-\epsilon_{inf}}{(\epsilon_{inf}-1)^2\tau^2}$, which gives the solution:
\be
\label{solpertinfT}
v=d^{inf}_1\sqrt{|\tau|}H_{\nu}(|k\tau|)~,~d^{inf}_2\sqrt{|\tau|}H_{-\nu}(|k\tau|)~,
\ee
and also
\be
\label{gammainf}
\gamma=\sqrt{2}\frac{v}{a}=d^{inf}_1\frac{2}{\sqrt{\pi}}Hk^\nu|\tau|^{\frac{3}{2}+\nu}~,~d^{inf}_2\frac{2}{\sqrt{\pi}}Hk^{-\nu}|\tau|^{\frac{3}{2}-\nu}~
\ee
in superhorizon region. Similar to the scalar case, the initial condition of $v$ can be obtained from Eq. (\ref{eompertinfT}) by considering the large $k$ region:
\be
v_i=\frac{1}{\sqrt{2k}}e^{ik\tau}~,
\ee
which can be combined with Eq. (\ref{solpertinfT}) to get: $d^{inf}_1=\sqrt{\pi}/2$. The power spectrum and spectral index of the tensor perturbation are then
\bea
\label{spectruminfT}
P^{T}_{inf}&\equiv&4\times\frac{k^3}{2\pi^2}|\gamma|^2=\frac{2H^2}{\pi^2}|k\tau|^{3+2\nu}~,\\
n^{T}_{inf}&\equiv&\frac{d\ln P^{T}_{inf}}{d\ln k}\simeq -2\epsilon_{inf}~,
\eea
from which we can see that the tensor perturbation is also nearly scale-invariant. Moreover, from Eq. (\ref{spectruminfS}) and Eq. (\ref{spectruminfT}) one gets the tensor/scalar ratio:
\be
\label{rinf}
r_{inf}\equiv\frac{P^{T}_{inf}}{P^{S}_{inf}}=16\epsilon_{inf}~,
\ee
which is within the current constraint: $r<0.07$ ($95\%$ C.L.) \cite{Akrami:2018odb} if $\epsilon$ is less than $4\times 10^{-3}$.

\subsection{matter contraction}
Besides inflation, there are also alternative possibilities on evolution of the early universe which have been deeply discussed in the literature, e.g. the universe may have experienced contraction before its expansion. If there is a smooth transition from contraction to expansion, it will be a nice way to avoid  the Big-Bang singularity and explain the origin of our universe. Here we consider the matter bounce model \cite{Finelli:2001sr, Wands:1998yp}, where the universe begins with a contraction phase with evolution behavior like ordinary matter, namely $w\simeq 0$. Different from inflation case where the perturbations are generated in expanding phase and are dominated by the constant mode, the perturbations in matter bounce models can be generated in contracting phase, therefore the varying mode becomes growing and then dominate over the constant one. In the following we will see that, if the universe contracts with $w\simeq 0$, one can also get scale-invariant scalar and tensor power spectrum.

The action of the universe in contracting phase is
\be
\label{actionmc}
S_{mc}=\int d^4x\sqrt{-g}\left[\frac{M_{Pl}^2R}{2}-\frac{1}{2}(\nabla\phi)^2-V(\phi)\right]~,
\ee
and different from the inflation case, in order to get matter-like contraction, we don't have slow-roll condition any longer. Instead, one needs the (average) value of kinetic and potential terms be nearly the same, $\langle\dot\phi^2\rangle\simeq\langle2V(\phi)\rangle$, such that $\langle w\rangle \simeq \langle p \rangle/\rho \simeq 0$. A possible choice to realize so is to have a concave shape, such as $V(\phi)=m^2\phi^2/2$ where $m$ is the mass of the field, and the field oscillates around its minimum.

The scale factor in matter-contracting phase can be parameterized as:
\be
a\sim\tau^{1/(\epsilon_{mc}-1)}~,~~~\epsilon_{mc}\equiv\frac{3}{2}(1+w_{mc})\simeq \frac{3}{2}~,
\ee
and the scalar perturbation action up to the second order is the same as Eq. (\ref{actionpertinfS}), with $\epsilon_{inf}\rightarrow\epsilon_{mc}$. Therefore, the curvature perturbation $\zeta$ has a solution:
\be
\label{zetamc}
\zeta=\frac{c^{mc}_1H}{\sqrt{\pi\epsilon_{mc}}}k^\nu|\tau|^{\frac{3}{2}+\nu}~,~\frac{c^{mc}_2H}{\sqrt{\pi\epsilon_{mc}}}k^{-\nu}|\tau|^{\frac{3}{2}-\nu}~,
\ee
with $\nu=\frac{1}{\epsilon_{mc}-1}-\frac{1}{2}$. Since for $\epsilon_{mc}\simeq 3/2$, $\nu\simeq 3/2$, and from Eq. (\ref{zetamc}) one can see that first mode is constant while the second mode is growing \footnote{Note that here $H$ scales as $\tau^{-3}$.}. So the spectrum and spectral index will be determined by the second mode:
\bea
\label{spectrummcS}
P^{S}_{mc}&\equiv&\frac{k^3}{2\pi^2}|\zeta|^2=\frac{H^2}{8\pi^2\epsilon_{mc}}|k\tau|^{3-2\nu}~,\\
n^{S}_{mc}&\equiv&1+\frac{d\ln P^{S}_{mc}}{d\ln k}=5-\frac{2}{\epsilon_{mc}-1}~,
\eea
which is nearly scale-invariant as $\epsilon_{mc}$ is close to $3/2$.

Moreover, the tensor perturbation up to second order is the same as Eq. (\ref{actioninfT}), and the solution is given by
\be
\label{gammamc}
\gamma=d^{mc}_1\frac{2}{\sqrt{\pi}}Hk^\nu|\tau|^{\frac{3}{2}+\nu}~,~d^{mc}_2\frac{2}{\sqrt{\pi}}Hk^{-\nu}|\tau|^{\frac{3}{2}-\nu}~.
\ee

Consider $\nu\simeq 3/2$, the second mode (growing) is dominant, with $d^{mc}_2=\sqrt{\pi}/2$ by matching with the initial condition. The tensor spectrum and spectral index determined by the growing mode are
\bea
\label{spectrummcT}
P^{T}_{mc}&\equiv&4\times\frac{k^3}{2\pi^2}|\gamma|^2=\frac{2H^2}{\pi^2\epsilon_{mc}}|k\tau|^{3-2\nu}~,\\
n^{T}_{mc}&\equiv&\frac{d\ln P^{T}_{mc}}{d\ln k}=5-\frac{2}{\epsilon_{mc}-1}~.
\eea

From Eq. (\ref{spectrummcS}) and Eq. (\ref{spectrummcT}), one can also get the tensor/scalar ratio:
\be
r_{mc}\equiv\frac{P^{T}_{mc}}{P^{S}_{mc}}=16\epsilon_{mc}~,
\ee
which is too large although, provided $\epsilon_{mc}\simeq 3/2$. This is actually a severe problem of matter contraction models realized by a single scalar field. Actually for more complicated models, one can have non-trivial sound speed squared $c_s^2$, such that $P^{S}$ can be suppressed (or $r$ be raised) by a factor $c_s$. For small $c_s$, such as $c_s\sim 10^{-2}$, $r$ can be suppressed to allowed value by observations, however, as has been proved in \cite{Quintin:2015rta} as a no-go theorem, it will cause large non-Gaussianities which is again conflict with the data constraints. Another choice is to introduce more than one scalar field, so the scalar power spectrum will roughly be multiplied by a factor of the number of the fields $N$, while the tensor/scalar ratio is divided by the same $N$. For large enough $N$, $r$ can also be dropped into the allowed value.
\subsection{slow contraction}
Other than matter-like, the contraction could also be slow (however the energy density grows fast), in order to avoid cosmic anisotropies from dominating the universe \cite{Kunze:1999xp}. One of the most famous scenarios of contracting universe is the Ekpyrotic scenario proposed some decades ago by Steinhardt et al. \cite{Khoury:2001wf}, however, the original Ekpyrotic model can make neither scalar perturbation (adiabatic perturbation $\zeta$) nor tensor perturbation scale-invariant, therefore several mechanisms are carried out \cite{Khoury:2009my, Finelli:2002we, Hinterbichler:2011qk} to amend the model. In this section, we consider the Ekpyrotic model with a scalar field {\it nonminimally coupled to Einstein gravity}, as has been proposed in \cite{Ijjas:2015zma} where it is dubbed as the ``anamorphic universe" model. The action is of the form:
\be
\label{scaction}
S_{sc}=\int d^4x\sqrt{-g}\left[\frac{1}{2}f(\phi)R-\frac{1}{2}k(\phi)(\nabla\phi)^2-V(\phi)\right]~,
\ee
where $f(\phi)$ and $k(\phi)$ are functions of $\phi$. By taking into account the non-minimal coupling, this model can be made such that the universe behaves like Ekpyrosis in Jordan frame while behaves like inflation/matter contraction in Einstein frame \cite{Ijjas:2015zma} (see also \cite{Wetterich:2013aca}). From the action, one can get the equation of motion for $\phi$:
\be
\label{sceom}
\ddot\phi+3H\dot\phi+\frac{k_{,\phi}\dot\phi^2}{k(\phi)}+\frac{V_{,\phi}}{k(\phi)}-\frac{f_{,\phi}}{2k(\phi)}R=0~,
\ee
and the energy density and pressure:
\bea
\label{scrho}
\rho&=&3M_{pl}^2H^2~\nonumber\\
&=&f^{-1}(\phi)\left(\frac{1}{2}k(\phi)\dot\phi^2+V(\phi)-3H\dot f(\phi)\right)~,\\
\label{scp}
p&=&-2M_{pl}^2\dot H-3M_{pl}^2H^2~\nonumber\\
&=&f^{-1}(\phi)\left(\frac{1}{2}k(\phi)\dot\phi^2-V(\phi)+\ddot f(\phi)+2H\dot f(\phi)\right)~,
\eea
and the equation of state of $\phi$ is defined as $w\equiv p/\rho$.

One can see from above that, in the usual Ekpyrotic model where $f(\phi)=1$, the slow-contraction ($w\geq 1$) requires $k(\phi)>0$, $V(\phi)<0$. However, for anamorphic case, it doesn't need to be so. Actually, in order to get an inflation in Einstein frame, $V(\phi)$ still need to be positive. Moreover, the conditions of no-ghost and inflation-like behavior in Einstein frame are given by the inequality \cite{Ijjas:2015zma}
\be
0<3+2k(\phi)\frac{f(\phi)}{f_{,\phi}^2}<\epsilon_E<1,
\ee
where $\epsilon_E$ is the slow-roll parameter in Einstein frame. The three inequality symbols from left to right are required by i) the no-ghost condition; ii) contraction and iii) inflation in Einstein frame, respectively \cite{Ijjas:2015zma}. For $f(\phi)>0$, this condition requires $k(\phi)<0$.

Defining the slow-varying parameters:
\be
\delta_{sc}\equiv\frac{\dot f}{Hf}~,~\epsilon_{sc}\equiv-\frac{\dot H}{H^2}=\frac{k\dot\phi^2-H\dot f+\ddot f}{2H^2f}~,
\ee
one has
\be
\frac{k(\phi)\dot\phi^2}{H^2f}\simeq2\epsilon_{sc}+\delta_{sc}+\epsilon_{sc}\delta_{sc}-\delta_{sc}^2~,
\ee
where we ignored the time variation of $\delta_{sc}$. Therefore the perturbed action (scalar part) from Eq. (\ref{scaction}) reads:
\be
\label{actionscS}
\delta_2S_{sc}^{(S)}=\int d^4x a^3Q\left[\dot\zeta^2-\frac{c_s^2}{a^2}(\partial\zeta)^2\right]~,
\ee
where
\bea
Q&=&\frac{2f}{(2+\delta_f)^2}\left[3\delta_f^2+\frac{2k\dot\phi^2}{H^2f}\right]=f\bar{\epsilon}_{sc}~,\\
\bar{\epsilon}_{sc}&\equiv&\frac{\delta_{sc}+2\epsilon_{sc}}{2+\delta_{sc}}~,\\
c_s^2&=&1~.
\eea
The equation of motion can the same as Eq. (\ref{eompertinfS}), except that $z=a(t)\sqrt{2Q}$. For slow contracting we set $a\sim |\tau|^{1/(\epsilon_{sc}-1)}$, with $\epsilon_{sc}\geq 3$. Moreover, $f\sim a^{\delta_{sc}}\sim \tau^{\delta_{sc}/(\epsilon_{sc}-1)}$, therefore
\be\label{zpp}
\frac{z^{\prime\prime}}{z}\rightarrow \frac{(2+\delta_{sc})(4+\delta_{sc}-2\epsilon_{sc})}{4(\epsilon_{sc}-1)^2|\tau|^2}~,
\ee
and the solution is
\bea
\label{zetasc}
\zeta&=&\frac{u}{z}=\frac{c^{sc}_1H}{\sqrt{\pi Q}}k^\nu|\tau|^{\frac{3}{2}+\nu}~,~\frac{c^{sc}_2H}{\sqrt{\pi Q}}k^{-\nu}|\tau|^{\frac{3}{2}-\nu}~,\\
\nu&=&\frac{2+\delta_{sc}}{2(\epsilon_{sc}-1)}-\frac{1}{2}~,
\eea
in the superhorizon region. Moreover, the tensor part perturbation action is
\be
\label{actionscT}
\delta_2S^{(T)}_{sc}=\int d^4x \frac{a^3f}{8}\sum_{s=+,\times}\left[\dot\gamma^{(s)2}-\frac{(\partial\gamma^{(s)})^2}{a^2}\right]~,
\ee
and the equation of motion is the same as Eq. (\ref{eompertinfT}), except that $a^{\prime\prime}/a$ replaced by $(a\sqrt{f})^{\prime\prime}/(a\sqrt{f})$, and $v=a\gamma\sqrt{f/2}$. Taking into account Eq. (\ref{zpp}), the solution is
\be
\label{gammasc}
\gamma=\frac{v}{a}\sqrt{\frac{2}{f}}=d^{sc}_1\frac{2}{\sqrt{\pi f}}Hk^\nu|\tau|^{\frac{3}{2}+\nu}~,~d^{sc}_2\frac{2}{\sqrt{\pi f}}Hk^{-\nu}|\tau|^{\frac{3}{2}-\nu}~
\ee
in superhorizon region.

{\it \bf Conformally dual to inflation.} In this case $\nu\simeq -3/2$ (with the deviation of ${\cal O}(\epsilon_E)$) is required, therefore one has
\be
\label{conditionsc1}
\delta_{sc}\simeq -2\epsilon_{sc}+{\cal O}(\epsilon_E)(\epsilon_{sc}-1)\ll -1~.
\ee
Moreover, one also has $H/\sqrt{Q}\sim H/\sqrt{f}\sim|\tau|^{-(2\epsilon_{sc}+\delta_{sc})/[2(\epsilon_{sc}-1)]}\sim |\tau|^{{\cal O}(\epsilon_E)}$, which is slowly varying. Actually, this is nothing but $H$ in Einstein frame, which describes inflation. Therefore, one can see that alike the inflation case, the second mode of Eq. (\ref{zetasc}) is decaying and the first mode is constant and dominating. By matching with initial condition which is the same as Eq. (\ref{uini}), one can get $c^{sc}_1=\sqrt{\pi}/2$. So the power spectrum and the spectral index of the Ekpyrotic model, which are measured at the end of the Ekpyrotic phase, are
\bea
\label{spectrumscS1}
P^{S}_{sc}&\equiv&\frac{k^3}{2\pi^2}\Big|\frac{u}{z}\Big|^2=\frac{H_\ast^2}{8\pi^2f_\ast\bar{\epsilon}_{sc}}(|k\tau_\ast|)^{3+2\nu}~,\\
n^{S}_{sc}&\equiv&1+\frac{d\ln P^{S}_{sc}}{d\ln k}=3+\frac{2+\delta_{sc}}{\epsilon_{sc}-1}~,\\
\label{spectrumscT1}
P^{T}_{sc}&\equiv&4\times\frac{k^3}{2\pi^2}\Big|\frac{v}{a}\Big|^2=\frac{2H_\ast^2}{\pi^2f_\ast\bar{\epsilon}_{sc}}(|k\tau_\ast|)^{3+2\nu}~,\\
n^{T}_{sc}&\equiv&\frac{d\ln P^{T}_{sc}}{d\ln k}=2+\frac{2+\delta_{sc}}{\epsilon_{sc}-1}~,
\eea
which are nearly scale-invariant considering Eq. (\ref{conditionsc1}). Moreover, from Eq. (\ref{spectrumscS1}) and Eq. (\ref{spectrumscT1}) one gets the tensor/scalar ratio:
\be
r_{sc}\equiv\frac{P^{T}_{sc}}{P^{S}_{sc}}=16\bar{\epsilon}_{sc}\sim 16\epsilon_E~,
\ee
which is the same as the inflation case.

{\it \bf Conformally dual to matter-contraction.} In this case $\nu\simeq 3/2$ is required, therefore one has
\be
\label{conditionsc2}
\delta_{sc}\simeq 4\epsilon_{sc}-6 \gg 1~,
\ee
and also, $H/\sqrt{Q}\sim |\tau|^{-3}$, which is growing. Actually, this is nothing but $H$ in Einstein frame, which describes matter contraction, where the second mode of Eq. (\ref{zetasc}) is dominating. The power spectrum and spectral index are
\bea
\label{spectrumscS2}
P^{S}_{sc}&\equiv&\frac{k^3}{2\pi^2}\Big|\frac{u}{z}\Big|^2=\frac{H_\ast^2}{8\pi^2f_\ast\bar{\epsilon}_{sc}}(|k\tau_\ast|)^{3-2\nu}~,\\
n^{S}_{sc}&\equiv&1+\frac{d\ln P^{S}_{sc}}{d\ln k}=5-\frac{2+\delta_{sc}}{\epsilon_{sc}-1}~,\\
\label{spectrumscT2}
P^{T}_{sc}&\equiv&4\times\frac{k^3}{2\pi^2}\Big|\frac{v}{a}\Big|^2=\frac{2H_\ast^2}{\pi^2f_\ast\bar{\epsilon}_{sc}}(|k\tau_\ast|)^{3-2\nu}~,\\
n^{T}_{sc}&\equiv&\frac{d\ln P^{T}_{sc}}{d\ln k}=4-\frac{2+\delta_{sc}}{\epsilon_{sc}-1}~,
\eea
which are nearly scale-invariant considering Eq. (\ref{conditionsc2}). Moreover, one gets the tensor/scalar ratio:
\be
r_{sc}\equiv\frac{P^{T}_{sc}}{P^{S}_{sc}}=16\bar{\epsilon}_{sc}~,
\ee
where $\bar{\epsilon}_{sc}\simeq 3/2$ as the matter contraction case.

\subsection{slow expansion}
Another interesting example is to assume that the universe has experienced a slowly expansion period in the early universe \cite{Piao:2003ty}. This scenario can also be realized based on more fundamental theories, one of the examples being the ``string gas model" (see \cite{Brandenberger:2011et} for a review). Similar to the slow-contracting case, the evolution of the scale factor can be parameterized as
\be
a\sim |\tau|^{1/\epsilon_{se}-1}~,~~~\epsilon_{se}\ll -1~.
\ee
As an expansion scenario, one does not need to worry about the anisotropy problem. However, as in the slow-contracting case, the original slow-expansion model with a single scalar field cannot obtain scale-invariant power spectrum either \footnote{There are however some mechanism call ``adiabatic mechanism" to obtain scale-invariant perturbations by evolving the slow-varying parameter $\varepsilon$, see \cite{Piao:2010bi}. However, we will not discuss this situation in the current status.}. Therefore, one also need to nonminimally couple the field to gravity to make it behave like inflation in its Einstein frame, so that the perturbations also behave like those of inflation \cite{Piao:2011mq}.

The action can be written the same as that of the slow contraction in Eq. (\ref{scaction}), namely,
\be
\label{seaction}
S_{se}=\int d^4x\sqrt{-g}\left[\frac{1}{2}f(\phi)R-\frac{1}{2}k(\phi)(\nabla\phi)^2-V(\phi)\right]~,
\ee
and the equation of motion, energy density and pressure can have the same expression as in Eq. (\ref{sceom}), Eq. (\ref{scrho}) and Eq. (\ref{scp}). Note that in minimal coupling case where $f(\phi)=1$, the slow-expansion ($w\ll-1$) requires $k(\phi)<0$, $V(\phi)>0$. However, in general case, the constraint on $k(\phi)$ can be loosened. Moreover, since the actions of scalar and tensor parts of perturbations, which are obtained by perturbing action Eq. (\ref{seaction}), have the same forms as those of Eq. (\ref{actionscS}) and Eq. (\ref{actionscT}), the solutions should also have the same form, namely:
\bea
\label{zetase}
\zeta&=&\frac{c^{se}_1H}{\sqrt{\pi Q}}k^\nu|\tau|^{\frac{3}{2}+\nu}~,~\frac{c^{se}_2H}{\sqrt{\pi Q}}k^{-\nu}|\tau|^{\frac{3}{2}-\nu}~,\\
\label{gammase}
\gamma&=&d^{se}_1\frac{2}{\sqrt{\pi f}}Hk^\nu|\tau|^{\frac{3}{2}+\nu}~,~d^{se}_2\frac{2}{\sqrt{\pi f}}Hk^{-\nu}|\tau|^{\frac{3}{2}-\nu}~,
\eea
with $\nu=\frac{2+\delta_{se}}{2(\epsilon_{se}-1)}-\frac{1}{2}$.

{\bf Conformally dual to inflation.} For $f\sim a^{\delta_{se}}\sim \tau^{\delta_{se}/(\epsilon_{se}-1)}$, one needs
\be
\label{conditionse1}
\delta_{se}\simeq -2\epsilon_{se}+{\cal O}(\epsilon_E)(\epsilon_{se}-1)\gg 1~,
\ee
to make $\nu\sim -3/2$. Therefore, the power spectrum and the spectral index of the slow-expansion model, which are measured at the end of the slow-expansion phase, are
\bea
\label{spectrumseS1}
P^{S}_{se}&\equiv&\frac{k^3}{2\pi^2}\Big|\frac{u}{z}\Big|^2=\frac{H_\ast^2}{8\pi^2f_\ast\bar{\epsilon}_{se}}(|k\tau_\ast|)^{3+2\nu}~,\\
n^{S}_{se}&\equiv&1+\frac{d\ln P^{S}_{se}}{d\ln k}=3+\frac{2+\delta_{se}}{\epsilon_{se}-1}~,\\
\label{spectrumseT1}
P^{T}_{se}&\equiv&4\times\frac{k^3}{2\pi^2}\Big|\frac{v}{a}\Big|^2=\frac{2H_\ast^2}{\pi^2f_\ast\bar{\epsilon}_{se}}(|k\tau_\ast|)^{3+2\nu}~,\\
n^{T}_{se}&\equiv&\frac{d\ln P^{T}_{se}}{d\ln k}=3+\frac{2+\delta_{se}}{\epsilon_{se}-1}~,
\eea
which are nearly scale-invariant considering Eq. (\ref{conditionse1}), where $\bar{\epsilon}_{se}\equiv(\delta_{se}+2\epsilon_{se})/(2+\delta_{se})$. Moreover, the tensor/scalar ratio is:
\be
r_{se}\equiv\frac{P^{T}_{se}}{P^{S}_{se}}=16\bar{\epsilon}_{se}\sim 16\epsilon_E~.
\ee

{\bf Conformally dual to matter-contraction.}
In like manner, one needs
\be
\label{conditionse2}
\delta_{se}\simeq 4\epsilon_{se}-6 \ll -1~,
\ee
to make $\nu\sim 3/2$. The power spectrum and spectral index is
\bea
\label{spectrumseS}
P^{S}_{se}&\equiv&\frac{k^3}{2\pi^2}\Big|\frac{u}{z}\Big|^2=\frac{H_\ast^2}{8\pi^2f_\ast\bar{\epsilon}_{se}}(|k\tau_\ast|)^{3-2\nu}~,\\
n^{S}_{se}&\equiv&1+\frac{d\ln P^{S}_{se}}{d\ln k}=5-\frac{2+\delta_{se}}{\epsilon_{se}-1}~,\\
\label{spectrumseT}
P^{T}_{se}&\equiv&4\times\frac{k^3}{2\pi^2}\Big|\frac{v}{a}\Big|^2=\frac{2H_\ast^2}{\pi^2f_\ast\bar{\epsilon}_{se}}(|k\tau_\ast|)^{3-2\nu}~,\\
n^{T}_{se}&\equiv&\frac{d\ln P^{T}_{se}}{d\ln k}=4-\frac{2+\delta_{se}}{\epsilon_{se}-1}~,
\eea
which is nearly scale-invariant considering Eq. (\ref{conditionsc2}), where $\bar{\epsilon}_{se}\equiv(\delta_{se}+2\epsilon_{se})/(2+\delta_{se})\simeq 3/2$. Moreover, the tensor/scalar ratio is:
\be
r_{se}\equiv\frac{P^{T}_{se}}{P^{S}_{se}}=16\bar{\epsilon}_{se}~,
\ee
where $\bar{\epsilon}_{se}\simeq 3/2$ as the matter contraction case.

\section{breaking the degeneracy: frame-invaraint variables}
\subsection{parameterization}
In the above, we have shown that many scenarios can have degenerate results of observables related to the 2 point correlation functions, such as the amplitude of power spectrum (both scalar and tensor), the spectral index and the tensor/scalar ratio. In order to make it clear, we first classify those scenarios into two groups, which we call as the {\it I-group} (inflation, and the slow contraction/expansion models that are conformally dual to it), and the {\it M-group} (matter contraction, and the slow contraction/expansion models that are conformally dual to it). From the above analysis, the two groups can give totally different tensor/scalar ratios and thus it is very easy to distinguish them by looking at the constraint of tensor/scalar ratio, however, for scenarios inside each group, it seems that these observables are totally degenerate, and it might be unlikely to use them, as we usually used, to distinguish these scenarios.

But are there any other methods/quantities that we can use to distinguish these ``seemingly degenerate" scenarios? In \cite{Ijjas:2015zma}, it has been suggested that the answer may be yes. According to \cite{Ijjas:2015zma}, one can define the frame-invariant variables:
\bea
\label{scalefactor}
&&\alpha_m\equiv a\frac{m}{M^{0}_{Pl}}~,~\alpha_{Pl}\equiv a\frac{M_{Pl}}{M^{0}_{Pl}}~,\\
\label{Hubble}
&&\Theta_m\equiv \frac{1}{M_{Pl}}\left(H+\frac{\dot m}{m}\right)~,~\Theta_{Pl}\equiv\frac{1}{M_{Pl}}\left(H+\frac{\dot{M}_{Pl}}{M_{Pl}}\right)~,
\eea
where $a$ is the FRW scale factor and $H\equiv \dot a/a$ is the Hubble parameter, both are frame-dependent, however. $m$ and $M_{Pl}$ are the particle mass and the Planck mass, respectively, and $M^{0}_{Pl}$ is a normalization factor. In the trivial case of General Relativity (GR), both $m$ and $M_{Pl}$ are constant. However, in the nonminimal coupling case, $M_{Pl}$ is a time-dependent function and $m$ remains constant in Jordan frame, and vice versa in Einstein frame. Therefore, $\alpha_m$ and $\Theta_mM_{Pl}$ represent the scale factor and Hubble parameter in Jordan frame, and $\alpha_{Pl}$ and $\Theta_{Pl}M_{Pl}$ represent those in Einstein frame, respectively.

Moreover, one can define the effective slow-roll parameters for $\Theta_m$ and $\Theta_{Pl}$ as:
\be
\label{slowroll}
\epsilon_m\equiv-\frac{d\ln(\Theta_mM_{Pl}/m)}{d\ln\alpha_m}~,~~~\epsilon_{Pl}\equiv-\frac{d\ln\Theta_{Pl}}{d\ln\alpha_{Pl}}~,
\ee
which are also frame-invariant variables. According to these definitions, $\epsilon_m$ ($\epsilon_{Pl}$) can go back to the usual definition: $\epsilon\equiv-\dot H/H^2$ in Jordan (Einstein) frame, respectively. More detailed analysis of such frame-invariant variables are presented in Appendix \ref{app}.

We're considering various scenarios whose perturbations can be conformally dual to inflation/matter-contraction in their Einstein frames, where there is no nonminimal coupling. When dual to inflation/matter-contraction, although $\Theta_{Pl}M_{Pl}$ is larger/smaller than zero, one can still have $\Theta_mM_{Pl}> 0$ for expanding background (inflation, slow expansion) and $\Theta_mM_{Pl}<0$ for contracting background (matter-contraction, slow contraction). Moreover, as demonstrated above, the $\epsilon_m$ (and also $\epsilon_{Pl}$) are totally different in the four scenarios. So using these four variables, the four scenarios can be distinguished, which has been shown in the Table \ref{table} below.
\begin{widetext}
\begin{table*}
\centering
\begin{tabular}{c|c|c|c|c|c|c}
\hline
&\multirow{2}{*}{\rm{Inflation}} & \multirow{2}{*}{\rm{Matter-contraction}} & \multicolumn{2}{c|}{\rm{slow contraction[JF]}} & \multicolumn{2}{c}{\rm{slow expanstion[JF]}} \\
\cline{4-7}
& & & \rm{Inflation[EF]} & \rm{MC[EF]} & \rm{inflation[EF]}  & \rm{MC[EF]} \\
\hline
$\Theta_mM_{Pl}$   &     $>0$        &        $<0$         &    $<0$      &    $ <0$    &   $>0$    &   $>0$   \\
\hline
$\epsilon_m$           &  $\simeq 0$  &  $\simeq 3/2$  &  $\geq 3$  &  $\geq 3$  &  $\ll -1$  &  $\ll -1$  \\
\hline
$\Theta_{Pl}M_{Pl}$ &    $>0$        &        $<0$         &    $>0$      &    $<0$     &   $>0$    &   $ <0$    \\
\hline
$\epsilon_{Pl}$ & $\simeq 0$ &  $\simeq 3/2$   &  $\simeq 0$  &  $\simeq 3/2$  & $\simeq 0$  & $\simeq 3/2$ \\
\hline
\end{tabular}
\caption{$\Theta_mM_{Pl}$, $\epsilon_m$, $\Theta_{Pl}M_{Pl}$ and $\epsilon_{Pl}$ for different scenarios. For the second and third columns, we consider inflation and matter contraction in GR, where they are the same both in Jordan and Einstein frames. For the last two columns, we consider slow contraction and slow expansion in its Jordan frame, where they can be transformed into inflation or matter contraction in their Einstein frames, respectively.}\label{table}
\end{table*}
\end{widetext}

In the following, we will show how we distinguish these scenarios by using a parametrization way. According to Eq. (\ref{Hubble}), $\Theta_mM_{Pl}$ can be written as:
\bea
\label{thetam}
\Theta_mM_{Pl}&=&\Theta_{Pl}M_{Pl}+\frac{\dot m}{m}-\frac{\dot M_{Pl}}{M_{Pl}}~,\\
&=&\Theta_{Pl}M_{Pl}+M_{Pl}^0(\Delta_m-\Delta_{Pl})~,
\eea
where we have defined two dimensionless variables, $\Delta_m\equiv\dot m/(mM_{Pl}^0)$ and $\Delta_{Pl}\equiv\dot M_{Pl}/(M_{Pl}M_{Pl}^0)$, which describe the relative running of particle mass $m$ and Planck mass $M_{Pl}$, respectively. Moreover, from the definition of the slow-roll parameters Eq. (\ref{slowroll}), although a little bit complicated, one can obtain the relationship between $\epsilon_m$ and $\epsilon_{Pl}$:
\be
\label{epsilonm}
\int \epsilon_mmdt=\frac{m\int \epsilon_{Pl}M_{Pl}dt}{M_{Pl}-M_{Pl}^0(\Delta_{Pl}-\Delta_m)\int \epsilon_{Pl}M_{Pl}dt}~.
\ee
Taking derivatives of both sides, and note that $Mdt=M_{Pl}^0dt_E$ one can get
\bea
\label{epsilonm}
\epsilon_{m}&=&\frac{1}{\Big[1-(D/\gamma)\int\epsilon_{Pl}M_{Pl}^{0}dt_{E}\Big]^{2}}~\nonumber\\
&\times&\Bigg[\epsilon_{pl}-(D/\gamma)\int\epsilon_{Pl}M_{Pl}^{0}dt_{E}~\nonumber\\
&&+\frac{\dot{D}-\Delta_{m}D}{\gamma^{2}}(\int\epsilon_{Pl}M_{Pl}^{0}dt_{E})^{2}\Bigg]~,
\eea
where $D\equiv\Delta_M-\Delta_m$, and $\gamma\equiv M_{Pl}/M_{Pl}^0$.

The integral interval is constrained by e-folding number in Einstein-frame:
\be
\label{efolding}
N_E=\int_{t_{E}^{0}}^{t_{E}}\Theta_{Pl}M_{Pl}^{0}dt_{E}
\ee
which is also frame-invariant. For I-group, in order to solve the Big-Bang problems, $N_E$ should be around, say, 60. However, for M-group, since the Big-Bang problems are considered in other mechanisms, the constraint on $N_E$ mainly comes from observation, therefore can be much shorter than $60$. For example, Ref. \cite{Liddle:2003as} gives a minimum e-folding number of 18 by BBN constraint on non-standard cosmologies, while in Ref. \cite{Elizalde:2014uba} people only considered the e-folding number to be as few as 12.

Our strategy is very simple: for either I-group or M-group, the Einstein-frame quantities $\Theta_{Pl}$ and $\epsilon_{Pl}$ can be easily parametrized. Therefore, $\Delta_m$, $\Delta_{Pl}$ and $\gamma$ can be used to describe the difference of Hubble parameter and slow-roll parameters between two frames, and once they are known, one can derive $\Theta_m$ and $\epsilon_m$ from Eq. (\ref{thetam}) and Eq. (\ref{epsilonm}), and according to Table \ref{table}, the evolution of the universe can be known.

Actually, there have been many discussions around constraining the running of both particle mass and Planck mass in the literature. For example, Refs. \cite{Barrow:2005qf, Mota:2006fz, Scoccola:2008yf, Landau:2008re, Scoccola:2009iz} discussed about constraints on variation of proton and electron masses from various ages to today, while Refs. \cite{Zahn:2002rr, Avilez:2013dxa, Li:2015aug, Ooba:2016slp, Alonso:2016suf} discussed about the change of Newtonian constant $G$ (corresponding to the Planck mass $M_{Pl}$) in scalar-tensor theories. However, these constraints seems only for current values within a few redshift, and as far as we know, the constraints, especially on the {\it time derivative} of $m$ and $M_{Pl}$, on very high redshifts like primordial age, seems still be lacking. Nevertheless, it is reasonable to expect that we can also obtain the constraints on early values of $\Delta_m$ and $\Delta_{Pl}$ in the future, and once we obtain those constraints, we can use them to determine what process our universe have experienced at that time.

However, even without constraints from observational data, one can still parametrize $\Delta_m$ and $\Delta_{Pl}$ to analyze their effects to cosmic evolution.
For either I-group or M-group where the scenarios behaves like inflation/matter-contraction in their Einstein frame, one can parametrize the Einstein frame scale factor and Hubble parameter as:
\be
\alpha_{Pl}=a_E\sim t_E^{1/\epsilon_E}~,~\Theta_{Pl}M_{Pl}^{0}=H_E=\frac{1}{\epsilon_Et_E}~,
\ee
with $\epsilon_E=\epsilon_{Pl}$ is a positive constant. Therefore from Eq. (\ref{efolding}) one has $N_E=\int_{t_{E}^{0}}^{t_{E}}dt_{E}/(\epsilon_{E}t_{E})$, which gives $t_{E}=t_{E}^{0}e^{\epsilon_{E}N_E}$. Choosing initial condition of $M_{Pl}^{0}t_{E}^{0}=1$ ($t_{E}^{0}$ is 1 Planck time), so $\int M_{Pl}^{0}dt_{E}=M_{Pl}^{0}t_{E}-M_{Pl}^{0}t_{E}^{0}=e^{\epsilon_{E}N_E}-1$, and from Eq. (\ref{epsilonm}) one has:
\bea
\label{para}
\Theta_{m}&=&\frac{1}{\epsilon_{E}}e^{-\epsilon_{E}N_E}-\frac{D}{\gamma}~,\nonumber\\
\epsilon_{m}&=&\frac{\epsilon_{E}}{\Big[1-(D/\gamma)\epsilon_{E}(e^{\epsilon_{E}N_E}-1)\Big]^{2}}~\nonumber\\
&\times&\Bigg[1-(D/\gamma)(e^{\epsilon_{E}N_E}-1)\nonumber\\
&&+\frac{\dot{D}/M_{Pl}^{0}-\Delta_{m}D}{\gamma^{2}}\epsilon_{E}(e^{\epsilon_{E}N_E}-1)^{2}\Bigg]~.
\eea

\subsection{dividing in parameter space}
From Eq. (\ref{para}), we can see that given the parameters in Einstein frame, namely $\epsilon_E$ and $N_E$, both quantities $\Theta_m$ and $\epsilon_m$ can be expressed in terms of some combinations of the parameters we defined, namely $D/\gamma$, $\dot D/(M_{Pl}^0\gamma^2)$, as well as $\Delta_m/\gamma$. Therefore, one can study how these quantities, required to satisfy various conditions in Table \ref{table}, are distributed in the parameter space formed by these parameters, as will be done in this section. Although the parameter space consists of 3 parameters and is 3 dimension, since in the following we only consider about scenarios in Jordan frame, the parameter space then consists of only 2 parameters (namely $D/\gamma$ and $\dot D/(M_{Pl}^0\gamma^2)$) and reduces to 2 dimension.

\begin{figure}[h]
\centering
\includegraphics[width=0.7\linewidth]{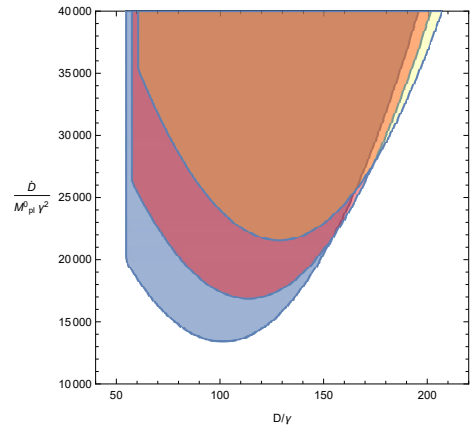}
\caption{The region in parameter space of $\{D/\gamma, \dot D/(M_{pl}^0\gamma^2)\}$ which corresponds to slow-contraction in Jordan frame while inflation in Einstein frame (I-group). The blue, red, yellow regions corresponds to $N_E=60,55,50$ respectively, with $\varepsilon_{E}=1.0\times 10^{-2}$.}
\label{I-g1}
\end{figure}

\begin{figure}[h]
 \centering
 \includegraphics[width=0.7\linewidth]{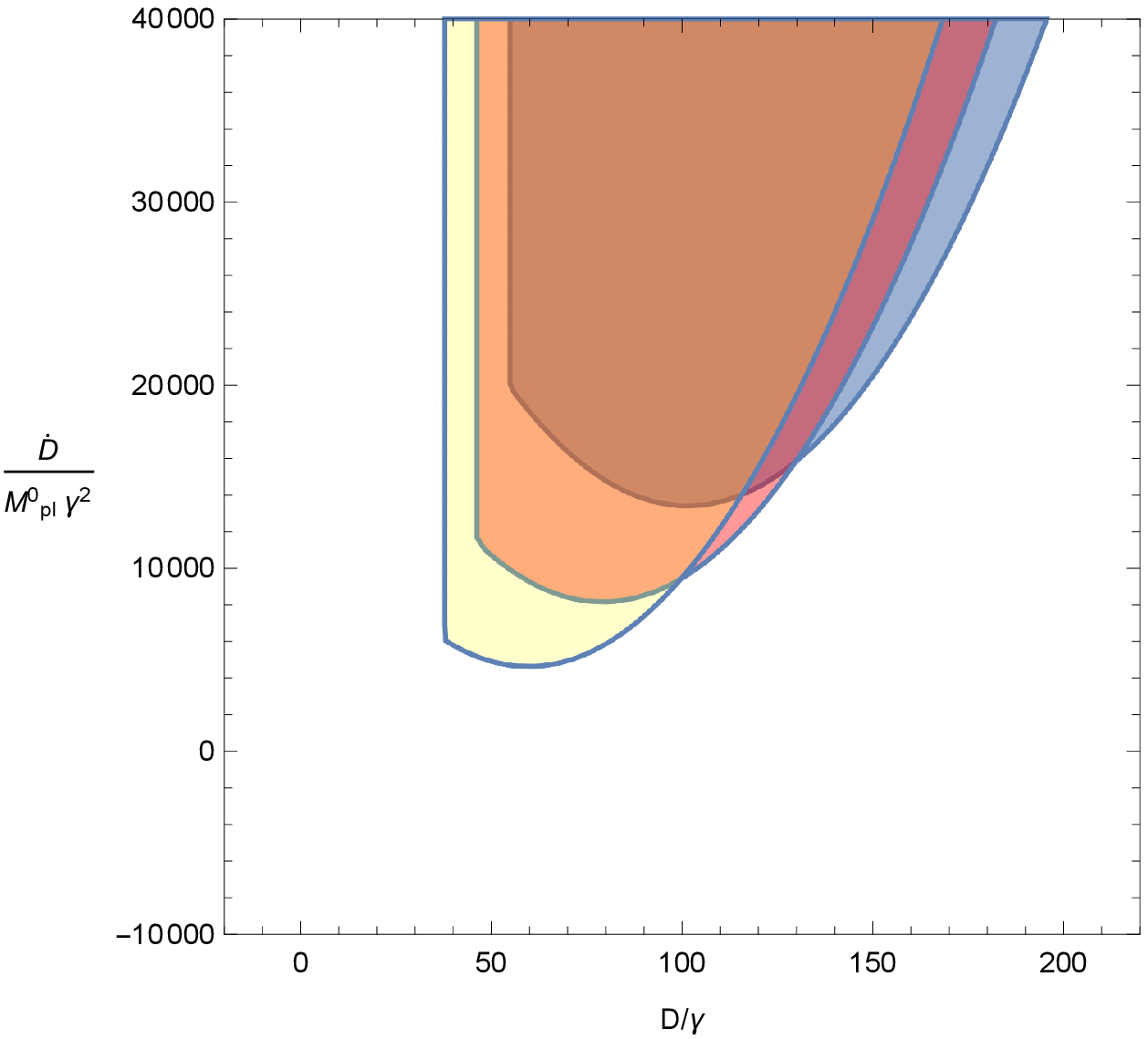}
 \caption{The region in parameter space of $\{D/\gamma, \dot D/(M_{pl}^0\gamma^2)\}$ which corresponds to slow-contraction in Jordan frame while inflation in Einstein frame (I-group). The blue, red, yellow regions corresponds to $\varepsilon_{E}=1.0\times 10^{-2}, 1.11\times 10^{-2},1.25\times 10^{-2}$ respectively, with $N_E=60$.}
 \label{I-g2}
\end{figure}

\begin{figure}[h]
 \centering
 \includegraphics[width=0.7\linewidth]{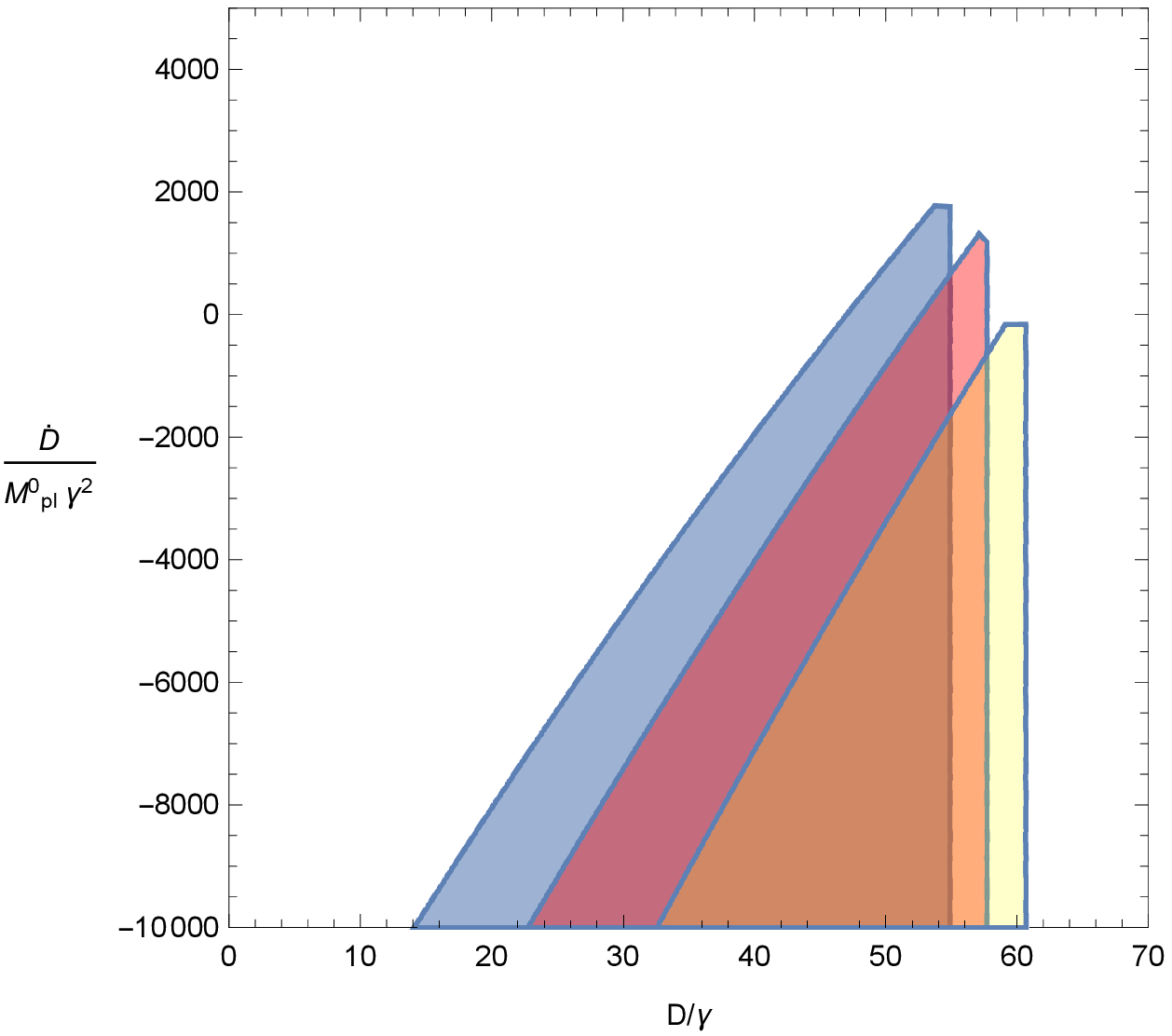}
 \caption{The region in parameter space of $\{D/\gamma, \dot D/(M_{pl}^0\gamma^2)\}$ which corresponds to slow-expansion in Jordan frame while inflation in Einstein frame (I-group). The blue, red, yellow regions corresponds to $N_E=60,55,50$ respectively, with $\varepsilon_{E}=1.0\times 10^{-2}$.}
 \label{I-g3}
\end{figure}

\begin{figure}[h]
 \centering
 \includegraphics[width=0.7\linewidth]{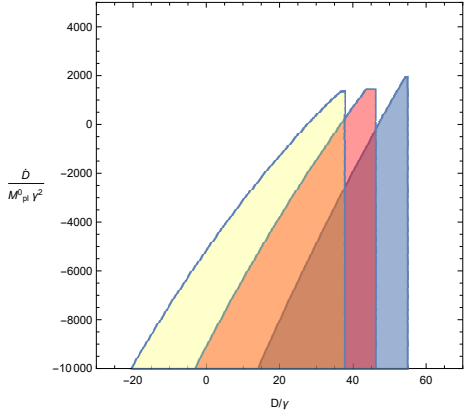}
 \caption{The region in parameter space of $\{D/\gamma, \dot D/(M_{pl}^0\gamma^2)\}$ which corresponds to slow-expansion in Jordan frame while inflation in Einstein frame (I-group). The blue, red, yellow regions corresponds to $\varepsilon_{E}=1.0\times 10^{-2}, 1.11\times 10^{-2},1.25\times 10^{-2}$ respectively, with $N_E=60$.}
 \label{I-g4}
\end{figure}

We consider the I-group and M-group separately. According to Table. \ref{table} and Eq. (\ref{para}), for I-group, we plot the region in the parameter space $\{D/\gamma, \dot D/(M_{Pl}^0\gamma^2)\}$ inside which the universe in Jordan frame evolves as slow-contraction and slow-expansion in Figs. \ref{I-g1}, \ref{I-g2}, \ref{I-g3} and \ref{I-g4},  respectively. In the plots, we choose various values of $\varepsilon_{E}$ and $N_E$, which are allowed by the current observations. The plots indicate that, although the information of perturbations of those scenarios are degenerated such that we cannot distinguish the scenarios in Jordan frame, as long as we can detect the parameters $D/\gamma$ and $\dot D/(M_{Pl}^0\gamma^2)$ and determine which region (shadowed or not) they are located, we can actually break the degeneracy. Moreover, as we see in the figure, in order to have slow-contraction or slow expansion, in general large $D/\gamma$ and/or $\dot D/(M_{Pl}^0\gamma^2)$ are needed, which indicates a large running of Planck mass in the early universe. This may be due to the fact that a small $\epsilon_E$ in Eq. (\ref{para}) will suppress the deviation of both $\Theta_m$ and $\epsilon_m$ from $\Theta_{Pl}$ and $\epsilon_{Pl}$, i.e. Hubble parameter and slow-roll parameter in Einstein frame.

Furthermore, for the case when $D=0$, the trivial case of GR will be recovered. In this case, the Einstein and Jordan frame coincide, with $\Theta_mM=\Theta_{Pl}=H$, $\epsilon_m=\epsilon_E$, and the universe is inflation even in Jordan frame.

\begin{figure}[h]
 \centering
 \includegraphics[width=0.7\linewidth]{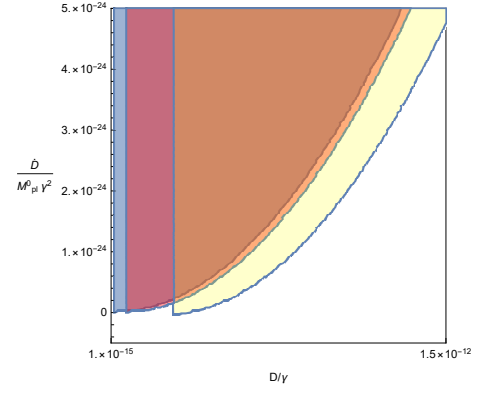}
 \caption{The region in parameter space of $\{D/\gamma, \dot D/(M_{pl}^0\gamma^2)\}$ which corresponds to slow-contraction in Jordan frame while matter-contraction in Einstein frame (M-group). The blue, red, yellow regions corresponds to $N_E=21,20,19$ respectively, with $\varepsilon_{E}=3/2$.}
 \label{M-g1}
\end{figure}

\begin{figure}[h]
 \centering
 \includegraphics[width=0.7\linewidth]{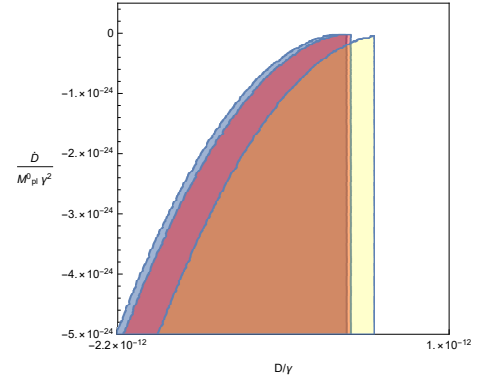}
 \caption{The region in parameter space of $\{D/\gamma, \dot D/(M_{pl}^0\gamma^2)\}$ which corresponds to slow-expansion in Jordan frame while matter-contraction in Einstein frame (M-group). The blue, red, yellow regions corresponds to $N_E=21,20,19$ respectively, with $\varepsilon_{E}=3/2$.}
 \label{M-g2}
\end{figure}

For M-group, we also plot the region in the parameter space $\{D/\gamma, \dot D/(M_{Pl}^0\gamma^2)\}$ inside which the universe in Jordan frame evolves as slow-contraction and slow-expansion in Figs. \ref{M-g1} and \ref{M-g2} respectively. In the plots, we choose several values of $N_E$ will fixing $\epsilon_E=3/2$. One could see from the figure that, in this case, even for small running of Planck mass, there could also be large possibility that we get slow evolution scenario in Jordan frame. Therefore it is easier to have diversity of scenarios, without the suppression effect of $\epsilon_E$. However, as we can also see from the figures, the results are a little bit sensitive to $N_E$ because of the exponential form. The difference of 1 e-fold will shift the shape by one or several order of magnitude in terms of $D/\gamma$.

Moreover, the case of $D=0$ also reduces to the GR case where the Einstein and Jordan frame coincide. Therefore, even in Jordan frame, the universe is matter-contraction. Note that all the above results are consistent with our previous works with specific examples \cite{Qiu:2012ia}.

\subsection{$f\left[\phi\left(t\right)\right]$: concrete examples}

The above subsection discussed about the probability of various cosmic scenarios in Jordan frame using the parameterization method. In order to give concrete examples, in this subsection we consider several models with different forms of $f\left[\phi\left(t\right)\right]$. Since $f$ is a function of $\phi$ and $\phi$ is a function of $t$, in this work we consider directly $f$ is a function of $t$. and for simplicity but without generality, we consider three cases,  including exponential form, power-law form and polynomial form.

According to conformal transformation, we have:
\be
dt_{E}=\sqrt{f\left(t_{J}\right)}dt_{J}~,~a_{E}=\sqrt{f\left(t_{J}\right)}a_{J}~,
\ee
which can give rise to:
\be
\dot{a_{J}}=\frac{da_{J}}{dt_{J}}=-\frac{1}{2}f^{-3/2}\dot{f}a_{E}+a_{E}^{\prime}~.
\ee
Actually, it can be easily seen that $\sqrt{f(t_J)}$ is equal to $M_{pl}/M^0_{pl}$. Therefore, we can rewrite Eq. (\ref{para}) as:
\bea
\label{thetammodel}
\Theta_{m}&=&\frac{1}{\varepsilon_{E}t_{E}}\left(1-\frac{\dot{f}\varepsilon_{E}t_{E}}{2f^{3/2}}\right)~,\\
\label{slowrollmodel}
\varepsilon_{m}&=&-f^{1/2}\frac{a_{J}}{\dot{a_{J}}}\left(\frac{\Theta_{m}^{\prime}}{\Theta_{m}}+\frac{\dot f}{2f^{3/2}}\right)~\nonumber\\
&=&-\frac{1}{\Theta_mM_{pl}^0}\left(\frac{\Theta_{m}^{\prime}}{\Theta_{m}}+\frac{\dot f}{2f^{3/2}}\right)~.
\eea

First of all, we consider $f(t_J)$ to be of exponential form, namely
\be
f\left(t_{J}\right)=\alpha e^{\beta t_{J}}~.
\ee
In this case, from Eqs. (\ref{thetammodel}) and (\ref{slowrollmodel}) we get
\bea
\Theta_{m}&=&\frac{1}{\varepsilon_{E}t_{E}}\left(1-\frac{\epsilon_Et_E}{t_E-C}\right)~,\\
\varepsilon_{m}&=&\frac{C(C-t_E)\epsilon_E}{M_{pl}^0[C-(1-\epsilon_E)t_E)]^2}~,
\eea
where $C=t_E^0-2\sqrt{\alpha}e^{\beta t_J^0}/\beta$ is the integral constant. Usually one has the freedom to choose the initial conditions $t_E^0=2\sqrt{\alpha}e^{\beta t_J^0}/\beta$ to make $C=0$, correspondingly giving $\Theta_{m}=(1-\epsilon_E)/\epsilon_Et_E$ and $\varepsilon_{m}=0$, which are independent of $\alpha$ and $\beta$. Since for I-group $t_{E}>0$, $0<\varepsilon_{E}\ll1$ while for M-group $t_{E}<0$, $\varepsilon_{E}\simeq 3/2$, one finds that in this case, no matter which group it is in, we have $\Theta_{m}>0$ and $\varepsilon_{m}=0$, therefore giving rise to inflation-like behavior in the Jordan frame.

However, there is a special case where $\beta\rightarrow 0$, which make the case reduce to GR case by letting $f(t)$ become a constant. In this case, $C$ will be very large unless a diverging $t_E^0$ is defined to cancel the divergence of $\beta$. Therefore from the equations above, it will give rise to $\Theta_m\rightarrow \Theta_E=1/\epsilon_E t_E$ and $\epsilon_m\rightarrow\epsilon_E$, as is expected.

Secondly, we consider $f(t_J)$ to be of power-law form, namely
\be
f\left(t_{J}\right)=\alpha t_{J}^{2\beta}~.
\ee
In this case, we get:
\bea
\Theta_{m}&=&\frac{1}{\varepsilon_{E}t_{E}}+\frac{\beta}{(\beta+1)(C-t_E)}~,\\
\varepsilon_{m}&=&\frac{\varepsilon_{E}[(\beta+1)(C-t_E)(C-t_E+C\beta)-\beta\epsilon_Et_E^2]}{M_{pl}^0[(\beta+1)(C-t_E)+\beta\epsilon_Et_E]^2}~,
\eea
where $C=t_E^0-\sqrt{\alpha}(t_J^0)^{\beta+1}/(\beta+1)$ is the integral constant. If we choose $t_E^0=\sqrt{\alpha}(t_J^0)^{\beta+1}/(\beta+1)$ to make $C=0$, the above formula will be reduced to
\be
\Theta_{m}\rightarrow\frac{1}{\varepsilon_{E}t_{E}}\left(1-\frac{\varepsilon_{E}\beta}{\beta+1}\right)~,
\varepsilon_{m}\rightarrow\frac{\varepsilon_{E}}{1+\beta\left(1-\varepsilon_{E}\right)}~,
\ee
which only depend on the parameter $\beta$.

For the case of I-group, we have $t_{E}>0$ and $0<\varepsilon_{E}\ll1$. Therefore, $\Theta_{m}>0$ for $\beta>-1$ or $\beta<-1/(1-\epsilon_E)$ while $\Theta_{m}<0$ for only narrow range of $-1/(1-\epsilon_E)<\beta<-1$. Moreover, for large value of $\beta$, we have $\epsilon\approx \epsilon_E$, which gives rise to inflation-like behavior in Jordan frame. The phase space for slow-contraction and slow expansion will be constrained to the regions of $(-1/(1-\epsilon_E),-(1-\epsilon_E/3)/(1-\epsilon_E)]$ and $[-(1+\epsilon_E)/(1-\epsilon_E),-1/(1-\epsilon_E))$, respectively.

For the case of M-group, we have $t_{E}<0$ and $\varepsilon_{E}=3/2$. Therefore $\Theta_{m}>0$ for $\beta>2$ or $\beta<-1$ while $\Theta_{m}<0$ for $-1<\beta<2$. Therefore for large value of $\beta$, we have $\epsilon\rightarrow 0$, which still gives rise to inflation-like behavior in Jordan frame. The phase space for slow-contraction and slow expansion will be constrained to the regions of $[1, 2)$ and $(2, 5)$, respectively. These results are consistent with the analysis in \cite{Qiu:2012ia}.

Finally, we consider $f(t_J)$ to be of polynomial form, namely
\be
f\left(t_{J}\right)=\left(\alpha+\beta t_{J}\right)^{2}~.
\ee
In this case, one has:
\bea
\Theta_{m}&=&\frac{1}{\varepsilon_{E}t_{E}}\left[1-\frac{\varepsilon_{E}t_{E}}{2(t_{E}+\rho-C)}\right]~,\\
\varepsilon_{m}&=&\frac{\varepsilon_{E}[2(t_{E}+\rho-C)(t_{E}+2\rho-2C)-\varepsilon_{E}t_{E}^{2}]}{M_{pl}^0[\varepsilon_{E}t_{E}-2(t_{E}+\rho-C)]^{2}}~.
\eea
where $\rho\equiv\alpha^2/(2\beta)$, and $C=t_E^0-\alpha t_J^0-\beta (t_J^0)^2/2$ is the integral constant. If we choose $t_E^0=\alpha t_J^0+\beta (t_J^0)^2/2$ to make $C=0$, the above formula will be reduced to
\bea
\Theta_{m}&=&\frac{1}{\varepsilon_{E}t_{E}}\left[1-\frac{\varepsilon_{E}t_{E}}{2(t_{E}+\rho)}\right]~,\\
\varepsilon_{m}&=&\frac{\varepsilon_{E}[2(t_{E}+\rho)(t_{E}+2)-\varepsilon_{E}t_{E}^{2}]}{M_{pl}^0[\varepsilon_{E}t_{E}-2(t_{E}+\rho)]^{2}}~,
\eea
which depend not only on the both $\alpha$ and $\beta$ (in terms of a single parameter $\rho$), but also on $t_E$ in a nontrivial manner.

For the case of I-group, we have $t_{E}>0$ and $0<\varepsilon_{E}\ll1$. Therefore, $\Theta_m>0$ for $\rho<-t_E$ or $\rho>(\varepsilon_{E}-2)t_E/2$, while $\Theta_m<0$ for $(\varepsilon_{E}-2)t_E/2<\rho<-t_E$. Moreover, when $\rho=\varepsilon_{E}t_E(\varepsilon_{E}+1)/[2(2\varepsilon_{E}-1)]-t_E$, we have $\varepsilon_{m}\approx\varepsilon_{E}$, which gives rise to inflation-like behavior in Jordan frame. Note that although we can make $\Theta_m>0$, there is no way to get slow-contraction phase, since the condition of $\epsilon_m>3$ cannot be satisfied. The phase space for slow-expansion will be constrained to the region of $(\varepsilon_{E}-2)t_E/2<\rho<(\varepsilon_{E}-2)t_E/[2(\varepsilon_{E}+1)]$.

For the case of M-group, we have $t_{E}<0$ and $\varepsilon_{E}=3/2$. Therefore, $\Theta_m>0$ for $-t_E/4<\rho<-t_E$, while $\Theta_m<0$ for $\rho<-t_E/4$  or $\rho>-t_E$. Moreover, when $\rho\approx\varepsilon_{E}t_E(\varepsilon_{E}+1)/[2(2\varepsilon_{E}-1)]-t_E\rightarrow-t_E/16$, we have $\varepsilon_{m}\approx3/2$, which gives rise to matter-contraction-like behavior in Jordan frame. The phase space for slow-contraction and slow-expansion will be constrained to the regions of $t_E/2<\rho<0$ and $-t_E/4<\rho<-t_E$, respectively.

\section{conclusion}
There are various possibilities of cosmological scenarios in the early universe. Although the observational data such as PLANCK can give more and more accurate constraints on those scenarios, there are still some that could not be distinguished from each other, not due to the precisions of observations, but due to the degeneracies in physics, caused by the dual relations such as conformal transformations. In order to distinguish those conformally related scenarios, we may need more variables to break the degeneracy.

In this paper, we made use of the frame-invariant variables defined in \cite{Ijjas:2015zma}, to discuss how we can distinguish those scenarios. These different values of frame-invariant variables can describe different scenarios according to Table. \ref{table}. These variables are constructed from the variation of Planck mass as well as particle mass. Although there are still no observational data to give constraints on these variables, we obtained various regions in parameter space corresponding to different scenarios. We found that to get slow-evolutions in I-group, one in general need large running of Planck mass, due to the suppressing effect of the slow-roll parameters. On the other hand, the requirement will be relaxed in M-group, but due to the exponential dependence of $\epsilon_m$ on $N_E$, the region of parameter space becomes more sensitive to $N_E$ than those in I-group.

Moreover, we also used several explicit models of the coupling function $f(\phi)$ with up to 2 free parameters, and got various constraints on those parameters to obtain different scenarios. Our results are consistent with previous analysis.

As a next step, it is therefore important to pursue whether and how can we really constraint those frame-invariant quantities using observational data, especially in terms of $\Delta_M$ and $\Delta_m$, since once we can get them constrained, we can localize where they are in the parameter space, and thus we can know which scenario the early universe would prefer. There have been some discussions on how to constrain those quantities for early time of the universe, such as the ``standard clock" approach \cite{Chen:2014joa, Chen:2015lza}. The further discussion about this project will be postponed to future research.

\begin{acknowledgments}
We thank Yi-Fu Cai, Xingang Chen, Taishi Katsuragawa, Shulei Ni, Yan Gong, Gong-Bo Zhao for helpful discussions. This work was supported by the National Natural Science Foundation of China with Grant No.~11875141 and No.~11653002, as well as the Fundamental Research Funds for the Central Universities with Grant No. CCNU19QN056.
\end{acknowledgments}

\appendix
\section{Frame-invariant variables}
\label{app}
In this appendix, we show how to construct variables that are independent of frame choice. A guiding principle is that, quantities that are frame-invariant should be either 1) quantities that are defined in a fixed (Jordan or Einstein) frame; 2) quantities that are equal in two (and also more) frames.

Let's start with the action in Jordan-frame:
\be
S_{J}=\int d^{4}x\sqrt{-g}\frac{M_{J}^{2}(t)}{2}R+\int m_{J}ds+\int d^{4}x\sqrt{-g}\mathcal{L}(\phi)~.
\ee
Using conformal transformation:
\bea
&&\tilde{g}_{\mu\nu}=\Omega^{2}g_{\mu\nu}~,~\tilde{a}=\Omega a~,~\tilde{dt}=\Omega dt~,~\tilde{ds}=\Omega ds~,\nonumber\\
&&\sqrt{-\tilde{g}}=\Omega^{4}\sqrt{-g}~,~\tilde{g}^{\mu\nu}=\Omega^{-2}g^{\mu\nu}~,~\tilde{R}=\Omega^{-2}R~,
\eea
One can transform into Einstein-frame:
\bea
S_{E}&=&\int d^{4}\tilde{x}\sqrt{-\tilde{g}}\Omega^{-2}\frac{M_{J}^{2}(t)}{2}\tilde{R}+\int m_{J}\Omega^{-1}\tilde{ds}\nonumber\\
&&+\int d^{4}\tilde{x}\sqrt{-\tilde{g}}\Omega^{-4}\mathcal{L}(\phi)~\nonumber\\
&=&\int d^{4}\tilde{x}\sqrt{-\tilde{g}}\frac{M_{E}^{2}}{2}\tilde{R}+\int m_E(t)\tilde{ds}+\int d^{4}\tilde{x}\sqrt{-\tilde{g}}\tilde{\mathcal{L}}(\tilde{\phi})\nonumber\\
\eea
Note that $M_{J}$ and $m_E$ are time-dependent, while $M_{E}$ and $m_{J}$ are time-independent. According to the above statement, $M_{J}$, $m_{E}$, $M_{E}$ and $m_{J}$ are actually all frame-invariant.

Moreover, from above we have:
\be
\Omega^{-1}M_{J}=M_{E}~,~\Omega^{-1}m_E=m_{J}~,
\ee
which means
\be
\label{omega}
\Omega=\frac{M_{J}}{M_{E}}=\frac{m_{J}}{m_{E}}~.
\ee
From this we can also have:
\be
\frac{m_{J}}{M_{J}}=\frac{m_{E}}{M_{E}}~,
\ee
so the quantity $m/M$ is frame-invariant.

Similarly, from conformal transformation we have
\be
\Omega=\frac{a_{E}}{a_{J}}=\frac{dt_{E}}{dt_{J}}~,
\ee
combined with (\ref{omega}), we can construct more frame-invariant variables.

For instance, we have
\be
\frac{a_{E}}{a_{J}}=\frac{m_{J}}{m_{E}}=\frac{M_{J}}{M_{E}}~,
\ee
so
\be
a_{E}m_{E}=a_{J}m_{J}~,~a_{E}M_{E}=a_{J}M_{J}~,
\ee
so both $am$ and $aM$ are frame-invariant. Moreover, we have:
\be
a_{J}=\frac{am}{m_{J}}~,~a_{E}=\frac{aM}{M_{E}}~.
\ee

In like manner, we have
\be
\frac{dt_{E}}{dt_{J}}=\frac{m_{J}}{m_{E}}=\frac{M_{J}}{M_{E}}~,
\ee
so
\be
m_{E}dt_{E}=m_{J}dt_{J}~,~M_{E}dt_{E}=M_{J}dt_{J}~,
\ee
so both $mdt$ and $Mdt$ are frame-invariant. Moreover, we have:
\be
dt_{J}=\frac{mdt}{m_{J}}~,~dt_{E}=\frac{Mdt}{M_{E}}~.
\ee

The Hubble parameter in Einstein and Jordan frame are
\bea
H_{E}&=&\frac{da_{E}}{a_{E}dt_{E}}=\frac{M_{E}}{M}(H+\frac{\dot{M}}{M})~,\nonumber\\
H_{J}&=&\frac{da_{J}}{a_{J}dt_{J}}=\frac{m_{J}}{m}(H+\frac{\dot{m}}{m})~,
\eea
where $H=\dot{a}/a$ and dot means derivative to $t$ in arbitrary frame.
Defining
\bea
\Theta_{Pl}&=&\frac{1}{M}(H+\frac{\dot{M}}{M})=\frac{H_{E}}{M_{E}}~,\nonumber\\
\Theta_{m}&=&\frac{1}{M}(H+\frac{\dot{m}}{m})=\frac{H_{J}}{m_{J}}\frac{m}{M}~,
\eea
both contains only frame-invariant variables. so both are frame-invariant.

One can also discuss furtherly about slow-roll parameters. The slow-roll parameter in Einstein and Jordan frames are
\be
\epsilon_{E}=-\frac{dH_{E}}{H_{E}^{2}dt_{E}}~,~\epsilon_{J}=-\frac{dH_{J}}{H_{J}^{2}dt_{J}}~.
\ee
Define frame-invariant variable
\be
\epsilon_{Pl}=-\frac{d\ln\Theta_{Pl}}{d\ln\alpha_{Pl}}~,~\epsilon_{m}=
-\frac{d\ln(\Theta_{m}M/m)}{d\ln\alpha_{m}}~,
\ee
and one can find that they are equal to $\epsilon_{E}$ and $\epsilon_{J}$ in each frame, respectively.

\end{document}